# Einstein's Pathway to the Equivalence Principle 1905-1907

Galina Weinstein

Between 1905 and 1907, Einstein first tried to extend the special theory of relativity in such a way so as to explain gravitational phenomena. This was the most natural and simplest path to be taken. These investigations did not fit in with Galileo's law of free fall. This law, which may also be formulated as the law of the equality of inertial and gravitational mass, was illuminating Einstein, and he suspected that in it must lie the key to a deeper understanding of inertia and gravitation. He imagined an observer freely falling from the roof of a house; for the observer there is during the fall – at least in his immediate vicinity – no gravitational field. If the observer lets go of any bodies, they remain relative to him, in a state of rest or uniform motion, regardless of their particular chemical and physical nature. The observer is therefore justified in interpreting his state as being "at rest". Newton realized that Galileo's law of free fall is connected with the equality of the inertial and gravitational mass; however, this connection was accidental. Einstein said that Galileo's law of free fall can be viewed as Newton's equality between inertial and gravitational mass, but for him the connection was not accidental. Einstein's 1907 breakthrough was to consider Galileo's law of free fall as a powerful argument in favor of expanding the principle of relativity to systems moving non-uniformly relative to each other. Einstein realized that he might be able to generalize the principle of relativity when guided by Galileo's law of free fall; for if one body fell differently from all others in the gravitational field, then with the help of this body an observer in free fall (with all other bodies) could find out that he was falling in a gravitational field.

## From the Special to the General Theory of Relativity

When did Einstein start to think about the problem of gravitation? Einstein published his first paper on the topic on **December 4, 1907**.[1] However, this date might be confusing. It would be better to say, the initial breakthrough regarding the equivalence principle occurred in 1907.

On June 20, 1933 Einstein went to Glasgow to give the University's first George Gibson Lecture. He spoke in English for twenty minutes in the University's Bute Hall on "The origins of the General Theory of Relativity." But the manuscript of the lecture was written in German, "Einiges über die entstehung der allgemeinen relativitätstheorie" ("something about the origin of the general theory of relativity").

---

[1] Einstein, Albert, "Über das Relativitätsprinzip und die aus demselben gezogenen Folgerungen", *Jahrbuch der Radioaktivität* 4, 1907, pp. 411-462.

Einstein's exposition was on the process which led him from the Special to the General Theory. Einstein explained his pathway towards the theory of gravitation *before* his 1907 breakthrough:[2]

"I came a step closer to the solution of the problem for the first time, when I attempted to treat the law of gravity within the framework of the special theory of relativity." Apparently, sometime **between September 1905 and September 1907** Einstein had already started to deal with the law of gravity within the framework of the special theory of relativity. When did he exactly start his work on the problem? Einstein did not mention any specific date, but he did describe the stages of his work presumably **prior to September 1907**.

At that time Einstein first tried to extend the special theory of relativity in such a way so as to explain gravitational phenomena. Actually this was the most natural and simplest path to be taken. And scientists at that time and even afterwards (in 1912) tried this path,

"Like most writers at the time, I tried to establish a *field-law* for gravitation, since it was no longer possible to introduce direct action at a distance, at least in any natural way, because of the abolition of the notion of absolute simultaneity.

The simplest thing was, of course, to retain the Laplacian scalar potential of gravity, and to complete the equation of Poisson in an obvious way by a term differentiated with respect to time in such a way, so that the special theory of relativity was satisfied. Also the law of motion of the mass point in a gravitational field had to be adapted to the special theory of relativity. The path here was less clearly marked out, since the inertial mass of a body could depend on the gravitational potential. In fact, this was to be expected on account of the inertia of energy".

But Einstein was not satisfied,[3]

"These investigations, however, led to a result which raised my strong suspicions. According to classical mechanics, the vertical acceleration of a body in the vertical gravitational field is independent of the horizontal component of its velocity. Here it depends together, the vertical acceleration of a mechanical system; respectively its center of gravity in such a gravitational field comes out independently of its internal kinetic energy. But according to the theory I tried, the acceleration of a falling body was not independent of its horizontal velocity, or the internal energy of the system.

This did not fit in with the old experiment that all bodies experience the same acceleration in a gravitational field. This law, which may also be formulated as the law of the equality of inertial and gravitational mass, was illuminating me, entering in now with its profound meaning. I was surprised in the highest degree at its existence and suspected that in it must lie the key to a deeper understanding of inertia and gravitation".

---

[2] Einstein, Albert, *The Origins of the General Theory of Relativity*, 1933, Glasgow Jackson: Wylie & co (a booklet of 12 pages); a translation of the German version, "Einiges über die Entstehung der allgemeinen Relativitätstheorie", and in Einstein, Albert**,** *Mein Weltbild*, 1934, Amsterdam: Querido Verlag, pp. 250-251; Einstein, Albert, *Ideas and Opinions*, 1954, New Jersey: Crown publishers (translated from Seelig's 1934 edition), pp. 286-287.
[3] Einstein, 1934, pp. 250-251; Einstein, 1954, pp. 286-287.

In a 1921 unpublished draft of a paper for *Nature* magazine, "Fundamental Ideas and Methods of the Theory of Relativity, Presented in Their Development", Einstein had explained in a more picturesque manner what he meant by "the key to a deeper understanding of inertia and gravitation":[4]

"When I (in Y. 1907) [in Bern] was busy with a comprehensive summary of my work on the special theory relativity for the 'Jahrbuch für Radioaktivität und Elektronik', I also had to try to modify Newton's theory of gravitation in such a way that its laws fitted into the theory. Attempts in this direction showed the feasibility of this enterprise, but did not satisfy me, because they had to be based upon unfounded physical hypotheses. Then there came to me the happiest thought of my life in the following form:

The gravitational field is considered in the same way and has only a relative existence like the electric field generated by magneto-electric induction. *Because for an observer freely falling from the roof of a house there is during the fall* – at least in his immediate vicinity – *no gravitational field*. Namely, if the observer lets go of any bodies, they remain relative to him, in a state of rest or uniform motion, regardless of their particular chemical and physical nature. The observer is therefore justified in interpreting his state as being 'at rest'.

The extremely strange experimental law that all bodies fall in the same gravitational field with the same acceleration, immediately receives through this idea a deep physical meaning. If there were just one single thing that fell differently in a gravitational field from the others, the observer could recognize with its help that he was in a gravitational field and that he was falling in the latter. But if such a thing does not exist – as experience has shown with great precision – then there is no objective reason for the observer to regard himself as falling in a gravitational field. Rather, he has the right to consider his state at rest with respect to gravitation, and his environment as field-free.

The experimental fact of independence of the material of acceleration, therefore, is a powerful argument for the extension of the relativity postulate to coordinate systems moving nonuniformly relative to each other".

---

[4] Einstein, Albert, "Grundgedanken und Methoden der Relativitätstheorie in ihrer Entwicklung dargestellt", 1921, Unpublished draft of a paper for *Nature* magazine, *The Collected Papers of Albert Einstein, Vol. 7: The Berlin Years: Writings, 1918–1921* (*CPAE*, Vol. 7), Janssen, Michel, Schulmann, Robert, Illy, Jószef, Lehner, Christoph, Buchwald, Diana Kormos (eds.), Princeton: Princeton University Press, 1998, p. 265.

Already in 1919, Einstein told a *New York Times* correspondent (who came to his Berlin home to interview him) about the above thought experiment. The correspondent reported about it in the following way, "It was during the development of the formulas for difform motions that the incident of the man falling from the roof gave me the idea that gravitation might be explained by difform motion". The correspondent transformed the thought experiment from Bern to Berlin and into a realistic amusing story. Presumably Einstein told the correspondent the story in this way, and he did not notice that Einstein was fooling him: "It was from his lofty library, in which this conversation took place, that he observed years ago a man dropping from a neighboring roof – luckily on a pile of soft rubbish – and escaping almost without injury. This man told Dr. Einstein that in falling he experienced no sensation commonly considered as the effect of gravity, which, according to Newton's theory, would pull him down violently toward the earth".[5]

Newton's apple is also mentioned in this article. Newton must have had something in mind when he compared the moon's centrifugal force with gravity – one of the hints leading him to the universal law of gravitation – and there is every reason to believe that the fall of an apple gave rise to it. William Stukely wrote that Newton told him the story and Conduitt also reported that Newton told him the same story about the apple; and others also reported that Newton was musing in his mother's garden in Lincolnshire, there came to him a thought about gravitation upon seeing a falling apple.[6]

However, it seems that *Einstein was not inspired by Newton's apple*. He was inspired by Galileo's law. Newton recognized that this law implied the equality of inertial and gravitational mass.

**Galileo's Principle**

Galileo's law of falling bodies is related with the story about the Leaning tower of Pisa. Most historians of science believe this is a legend, but it might not be so much a legend. Galileo began his lectures as professor of mathematics at the University of Pisa in 1589. During his term at Pisa Galileo revised and completed his treatise *De motu*, using material from earlier dialogue on motion and the intermediate version. It is probable that *De motu* was completed in Galileo's last year at Pisa, 1591-1592.[7] In 1591 Galileo examined in the *De motu* bodies of the same kind falling through the same medium. He wrote in chapter 6 of *De motu*, "In the case of bodies moving naturally, as in weights on a balance, the cause of all motions – up as well as down – can be referred to weight alone".[8] Perhaps, says Stillman Drake the known Galileo scholar, the best known story about Galileo relates to the Leaning Tower of Pisa.[9] And it might be dated to 1591-1592.

---

[5] "Einstein Expounds his New Theory", *New York Times* 1919, December 3.
[6] Westfall, Richard, S. *Never at Rest, A Biography of Isaac Newton*, 1980/1996, New York: Cambridge University Press, pp. 154-155.
[7] Drake, Stillman, *Galileo at Work: His Scientific Biography* (Chicago: University of Chicago Press, 1978), pp. 18-19.
[8] Drake, 1978, p. 22.

The exact words of its first appearance were in 1657 in a biography of Galileo by Vincenzo Viviani. Viviani was repeating his recollection of what Galileo himself had told him during his final years of blindness at Arcetri:[10]

"At this time, it appearing to him that for the investigation of natural effects there was necessary required a true knowledge of the nature of motion, there being a philosophical and popular axiom that 'Ignorance of motion is ignorance of nature', he quite gave himself over to its study; and then, to the great discomfort of all the philosophers, through experiences and sound demonstration and arguments, a great many conclusions of Aristotle himself on the subject of motion were shown by him to be false which up to that time had been held as most clear indubitable, as (among others) that speeds of unequal weights of the same material, moving through the same medium, did not at all preserve the ratio of their heaviness assigned to them by Aristotle, but rather, these all moved with equal speeds, he showing this by repeated experiments [*esperienze*] made from the height of the Leaning Tower of Pisa in the presence of other professors and all the students".

Drake says, "In fact what Viviani described was not an experiment at all; it was a demonstration. Galileo already knew what would happen and used the Leaning Tower to demonstrate this to others. That would have been in keeping with what is known about his flair for the dramatic". Galileo's demonstration caused difficulties to Aristotle's theory. Drake adds that Viviani stated clearly that the weights were of the same material and of different weights.[11]

Later in prison, in 1638 Galileo came back to the problem of natural motion of bodies and falling bodies. He gave an interesting theoretical reasoning for the law of falling bodies in the *Dialogue Concerning Two New Sciences*. On the first day of the *Discorsi* Salviati says that it is "possible to probe clearly, by means of a short and conclusive argument, that a heavier body does not move more rapidly than a lighter one provided both bodies are of the same material and in short such as those mentioned by Aristotle".[12]

Did Aristotle mention these bodies? Salviati did not quite check this. It is said that Aristotle believed that heavy bodies move and fall faster than light ones. Salviati: "But tell me, Simplicio, whether you admit that each falling body acquires a definite speed fixed by nature, a velocity which cannot be increased or diminished except by the use of force or resistance". Simplicio agrees somewhat with Salviati, and the latter goes on, "If then we take two bodies whose natural speeds are different, it is clear that on uniting the two, the more rapid one will be partly retarded by the slower, and the slower will be somewhat hastened by the swifter." Simplicio agrees again with Salviati: "You are unquestionably right".[13] And this is now also Simplicio's supposition.

---

[9] Drake, 1978, p. 19.
[10] Drake, 1978, pp. 19-20.
[11] Drake, 1978, p. 20.
[12] Galileo, Galilei, *Dialogues Concerning Two New Sciences*, translators, Henry Crew and Alfonso de Salvio, 1638/1914, New-York, Prometheus Books, 1991, section 107, p. 62.

Salviati now explains to Simplicio why "Aristotle's principle" leads to contradiction. Simplicio agrees that if we take two stones moving with different speeds, then on uniting them, the stone moving with a speed of eight will be retarded by the stone moving with a speed of four, and vice versa, the stone moving with the speed of four will be hastened by the other stone. Thus explains Salviati : "if a large stone moves with a speed of, say, eight while a smaller moves with a speed of four, then when they are united, the system will move with a speed less than eight; but the two stones when tied together make a stone larger than that which before moved with a speed of eight. Hence the heavier body moves with less speed than the lighter; the effect which is contrary to your supposition. Thus you see how, from your assumption that the heavier body moves more rapidly than the lighter one, I infer that the heavier body moves more slowly".[14]

Simplicio finds it difficult to understand. Salviati then imagines the two stones tied one to another (the smaller stone upon the larger), and one allows them to fall freely from some height. Then "during free and natural fall, the small stone does not press upon the larger and consequently does not increase its weight as it does when at rest". Simplicio then asks: "But what if we should place the larger stone upon the smaller?" Salviati explains that "Its weight would be increased if the larger stone moved more rapidly; but we have already concluded that when the small stone moves more slowly it retards to some extent the speed of the larger, so that the combination of the two, which is a heavier body than the larger of the two stones, would move less rapidly, a conclusion which is contrary to your hypothesis. We infer therefore that large and small bodies move with the same speed provided they are of the same specific gravity".[15]

Salviati extended his explanation of the two moving stones to falling stones. Consider again the two stones. The large stone that falls with speed eight and the small stone that falls with speed four. When tied together make a stone that weighs more than that which falls with a speed of eight, but which moves with a speed smaller than eight. Hence the heavier body falls with less speed than the lighter, and this leads to a contradiction. If a heavy stone and a light stone (of the same material) were let fall from the same height, the two would reach the ground exactly in the same time.

This is Galileo's law of falling bodies. Galileo's experience was later formulated in the following form: all bodies experience the same acceleration in a gravitational field.

---

[13] Galileo, 1638/1914, section 107, pp. 62-63.
[14] Galileo, 1638/1914, sections 107-108, pp. 62-63.
[15] Galileo, 1638/1914, sections 108-109, pp. 63-64.

In the opening paragraph of the *Principia*, Newton defines "the quantity of matter", and says it is proportional to the mass of the body. Newton says that "the quantity of matter" "is proportional to the weight, as I have found by experiments on pendulums".[16] Hence, approximately in 1685, Newton realized that there was an (empirical) equality between inertial and gravitational mass.

Max Jammer explains:[17] "Newton postulated a proportionality between *vis inertiae* [inertial mass] and another fundamental characteristic of a given body, its *quntitas materiae* [quantity of matter]". According to Book III of the *Principia*, the *quntitas materiae* determines the magnitude of gravitational attraction. Newton there writes that all bodies gravitate in proportion to the quantity of matter which they contain. Newton showed in a series of experiments he describes in Book III, Proposition 6, Theorem 6, of the Principia, that the *quntitas materiae* is proportional to gravity or weight for a given locality.

Indeed Newton there writes:[18] "It has been, now of a long time, observed by others, that all sorts of heavy bodies (allowance being made for the inequality of retardation which they suffer from a small power of resistance in the air) descend to the earth *from equal heights* in equal times; […] I tried the thing in gold, silver, lead, glass, sand, common salt, wood, water, and wheat. [he prepared equal weights of those nine diverse substances, and concluded …] And therefore the quantity of matter in the gold […] was to the quantity of matter in the wood as the action of the motive force upon all the gold to the action of the same upon all the wood; that is, as the weight of the one to the weight of the other: and the like happened in the other bodies".

But in Corollary 5, further ahead, Newton states:[19] "The power of gravity is of a different nature from the power of magnetism; for the magnetic attraction is not as the matter attracted. Some bodies are attracted more by the magnet; others less; most bodies not at all. The power of magnetism in one and the same body may be increased and diminished; and is sometimes far stronger, for the quantity of matter, than the power of gravity; and in receding from the magnet decreases not in the duplicate but almost in the triplicate proportion of the distance, as nearly as I could judge from some rude observations".

The force of magnetism is thus not proportional to the quantity of matter (mass) unlike the power of gravity. Although Newton realized that the experimental law, "all sorts of heavy bodies… descend to the earth *from equal heights* in equal times", is connected with the equality of the inertial and gravitational mass; this connection was accidental.

---

[16] Newton, Isaac, *The Principia. Mathematical Principles of Natural Philosophy*, translated by Andrew Motte, 1726/1995, New York: Prometheus Books, Book I, p. 9.
[17] Jammer, Max, *Concepts of Mass in Classical and Modern Physics*, 1961/1997, New-York: Dover, pp. 72-73.
[18] Newton, 1726/1995, Book III, p. 330.
[19] Newton, 1726/1995, Book III, p. 333.

**Einstein's 1907 breakthrough**

In 1921 Einstein wrote that Galileo's law of experience – the law of free fall – *can be viewed as* Newton's equality between inertial and gravitational mass, and he called this equality "hypothesis of equivalence" [Aequivalenzhypothese].[20] Hence, for Einstein this connection *is not accidental*. But one should bear in mind that Einstein's 1907-1912 "Principle of Equivalence" [Aequivalenzprinzip] *is not* this "Hypothesis of Equivalence", which signifies Galileo's-Newton's law of free fall.[21] Einstein's principle of equivalence is an extension of this hypothesis of equivalence.

Einstein's 1907 breakthrough was to consider Galileo's principle of free fall as a powerful argument in favor of expanding the principle of relativity to systems moving non uniformly relative to each other. Einstein realized that he might be able to generalize the principle of relativity when guided by Galileo's principle or law, according to which all bodies in gravitational field have the same fall; for if one body fell differently from all others in the gravitational field, then with the help of this body an observer in free fall (with all other bodies) could find out that he was falling in a gravitational field.[22]

**In September 1907**, the editor of the *Yearbook for Radioactivity and Electronics*, *Jahrbuch der Radioaktivität und Elektronik*, Johannes Stark, asked Einstein to write a review article on the theory of relativity. **On September, 25**, **1907** Einstein replied that he would be happy to "deliver the desired report", but he wished to know the date Stark would like to receive the paper.[23]

**On October 4, 1907**, Stark sent Einstein a letter thanking him for his willingness to write the review article, and he told him about two of Planck's major publications, Max Laue's paper from the *Annalen der Physik*, and Stark's own reply to Plank's work.[24] **On October 7, 1907**, Einstein wanted to know: what was the approximate date that Stark wished to receive the paper on the relativity principle?[25]

---

[20] Einstein, 1921, p. 24, *CPAE*, Vol. 7, Doc. 31, p. 268.
[21] In 1921 Einstein wrote that Mach later criticized Newton's mechanics; he was (after Newton) the first to vividly feel and clearly illuminate the epistemological weakness of classical mechanics. However, the natural equality between inertia and gravitation "remained hidden to Mach" [blieb Mach verborgen]. Einstein, 1921, p. 24, *CPAE*, Vol. 7, Doc. 31, p. 268.
[22] Einstein, 1921, p. 22, *CPAE*, Vol. 7, Doc. 31.
[23] Einstein to Stark, September 25, 1907, *The Collected Papers of Albert Einstein*, *Vol. 5: The Swiss Years: Correspondence, 1902–1914* (*CPAE*, Vol. 5), Klein, Martin J., Kox, A.J., and Schulmann, Robert (eds.), Princeton: Princeton University Press, 1993, Doc. 58.
Einstein told Stark, "Also, I must say that I am not able to acquaint myself with *everything* published on the subject, because the library is closed during my free time" (from the patent office). He was thus acquainted – in addition to his own works – with only four papers at that time (among them Lorentz's 1904 electron theory paper, and two papers by Planck, with whom Einstein was corresponding). Einstein told Stark that he did not know of any other theoretical work relevant to the subject (of special relativity). He closed his letter to Stark by saying, "You would therefore do me a great favor if you could bring to my attention other publications, if you know about such". Accordingly, Einstein was still sitting in the patent office and was unaware of important papers written by leading scholars.

**On November 1, 1907**, Einstein told Stark: "I have now finished the first part of the work for your *Jahrbuch*"; "I am working diligently on the second [part] in my, unfortunately rather scarce, free time". The first part dealt with special relativity. Einstein estimated that the whole paper would be 40 printed pages long, and he told Stark that he hoped he would send him the manuscript "by the end of this month".[26] **The paper was published on December 4, 1907**.

While writing this paper, Einstein suddenly arrived at a breakthrough, which boosted his research towards the General Theory of Relativity. Thus it appears that Einstein arrived at this breakthrough sometime *during November 1907*.

In the 1907 review article Einstein presented a new system, a coordinate-dependant theory, which reminded the 1905 "principle of relativity": guided by Galileo's principle, Einstein postulated the principle of equivalence, *Aequivalenzprinzip*, and with physical reference systems and measuring rods and clocks, he arrived at new results. When no complicated mathematics enters into the theory, the extension of the principle of relativity turned to be quite natural and simple.

Four years later, in June 1911, Einstein published a paper in Prague in the *Annalen der Physik* "Uber den Einfluβ der Schwerkraft auf die Ausbreitung des Lichtes" ("On the Influence of Gravitation on the Propagation of Light"). Einstein was guided by Galileo's law of free fall towards formulating a more mature equivalence principle. He still did not leave the comfortable framework of a coordinate-dependant theory: the physical frame of reference, the system of measuring rods and clocks that gave physical meaning for points in space-time. He wrote: "we arrive at a very satisfactory interpretation of this empirical law, if we suppose that the systems *K* and *K'* are physically exactly equivalent, i.e., if we assume that we may just as well define the system *K* as being found in space free from gravitational fields, for we must then consider *K* as uniformly accelerated".[27]

As long as we deal with mechanical processes and when Newton's mechanics holds good then *K* and *K'* are equivalent. What about all physical processes? Einstein wished to demonstrate that *K* and *K'* are equivalent with respect to all physical processes, and thus that the laws of nature with respect to *K* will be in agreement with those with respect to *K'*. "By assuming this, we obtain a principle which, if it is true, has great heuristic meaning. For we obtained by theoretical consideration of the processes which take place relatively to a uniformly accelerating reference system, information as to the course of processes in a homogeneous gravitational field".[28] Einstein extended his 1907 crude equivalence principle and formulated a heuristic guiding equivalence principle.

---

[24] Stark to Einstein, October 4, 1907, *CPAE*, Vol. 5, Doc. 60.
[25] Einstein to Stark, October 7, 1907, *CPAE*, Vol. 5, Doc. 61.
[26] Einstein to Stark, November 1, 1907, *CPAE*, Vol. 5, Doc. 63.
[27] Einstein, Albert, "Uber den Einfluβ der Schwerkraft auf die Ausbreitung des Lichtes", *Annalen der Physik* 35, 1911, pp. 898-908; p. 899.

In 1920, Einstein wrote a short list of "my most important scientific ideas". The final three items on the list are: [29]

1907 Basic idea for the theory of relativity

1912 Recognition of the non-Euclidean nature of the metric and its physical determination by gravitation

1915 Field equations of gravitation. Explanation of the perihelion motion of Mercury.

The "basic idea for the theory of relativity" relies on the hypothesis of equivalence and the principle of equivalence.

*I wish to thank Prof. John Stachel from the Center for Einstein Studies in Boston University for sitting with me for many hours discussing special relativity and its history.*

---

[28] Einstein, 1911, p. 900.
[29] Stachel, John, "The First-two Acts", in Stachel, John, *Einstein from 'B' to 'Z'*, 2002, Washington D.C.: Birkhauser, pp. 261-292; p. 261.